# ER model Partitioning: Towards Trustworthy Automated Systems Development

Dhammika Pieris[1]
Department of Human Resource Management
Faculty of Commerce and Management Studies
University of Kelaniya
Sri Lanka

M. C Wijegunesekera[2]
Department of Software Engineering
Faculty of Computing and Technology
University of Kelaniya
Sri Lanka

N. G. J Dias[3]
Department of Computer Systems Engineering
Faculty of Computing and Technology
University of Kelaniya
Sri Lanka

*Abstract*— In database development, a conceptual model is created, in the form of an Entity-relationship(ER) model, and transformed to a relational database schema (RDS) to create the database. However, some important information represented on the ER model may not be transformed and represented on the RDS. This situation causes a loss of information during the transformation process. With a view to preserving information, in our previous study, we standardized the transformation process as a one-to-one and onto mapping from the ER model to the RDS. For this purpose, we modified the ER model and the transformation algorithm resolving some deficiencies existed in them. Since the mapping was established using a few real-world cases as a basis and for verification purposes, a formal-proof is necessary to validate the work. Thus, the ongoing research aiming to create a proof will show how a given ER model can be partitioned into a unique set of segments and use it to represent the ER model itself. How the findings can be used to complete the proof in the future will also be explained. Significance of the research on automating database development, teaching conceptual modeling, and using formal methods will also be discussed.

## I. Introduction

The Entity-Relationship (ER) model[1, 2] is widely used to create conceptual schemas (conceptual models) to represent application domains in the field of Information Systems development. However, when an ER model is transformed to a Relational Database Schema (RDS) of the relational model, some critical information modeled on the ER model may not be represented meaningfully on the RDS [3-5]. This situation causes a loss of information during the transformation process[5, 6].

Min-max constraints, role names, composite attributes, subtype/supertype hierarchies, and certain relationship types are frequently lost in the transformation process[5][13].

Previous studies undertaken by other researchers on information loss [6-11] were of varying opinion. Some researches proposed ignoring the information that is lost during the transformation process and accepting only the information, that is, actually transformed. This proposal is called information reducing transformation (e.g., [8, 9].) Researches in [7] and [10] suggested that the min-max constraints that cannot be transformed and represented on the RDS to be directly implemented in the database system via triggers and stored procedures. This is a way of bypassing the RDS. According to [11], min-max constraints can be represented as a set of functions in a separate schema, external to the RDS. The functions are then implemented in the database as a program written in extended SQL (e.g., PL/SQL or T/SQL). The method is also a way of bypassing the RDS. The research in [6] indicated that supertype/subtype hierarchies that could be lost during a transformation could be directly implemented in the database system. It is also a way of bypassing the RDS. As indicated in [10] and [11], min-max constraints can be directly implemented in user application programs. It is a way of bypassing the logical level RDS as well as the physical level database.

In summary, some previous research suggests bypassing the logical level– that is, the RDS– and implementing the lost information directly on the physical level. Some others suggest bypassing both the logical level and the physical level and implementing the lost information directly in user application programs. Some other researchers proposed ignoring the information that is lost during the transformation process, suggesting that the information that is actually preserved is adequate.

However, in contrast to bypassing the RDS and ignoring the lost information, in our study, we focus on preserving information and representing them on the logical level RDS as much as possible.

According to [12], if the information is preserved when a conceptual schema (e.g., ER model) is transformed to a logical schema (e.g., RDS) (forward transformation), the logical schema should be able to reverse back to the conceptual schema (reverse transformation) by means of reverse applying the steps of the algorithm used for the forward transformation process. We based our research on this theory proposed by [12].

We argue that if the forward transformation can create a one-to-one and onto mapping from the ER model to the RDS, the RDS could be reversed back to the ER model. The RDS could be reversed back to the ER model means, according to [12], the information is preserved in the transformation process from the ER model to the RDS.

However, during our previous studies, we found that the deficiencies that exist in the ER model and the transformation algorithm hinder such a one-to-one and onto mapping is being established in the forward transformation process. [5, 13] [14-16]. We then modified the ER model and the transformation algorithm [5, 14, 15], eliminated the deficiencies and avoid that hindrance. Accordingly we established a one-to-one and onto mapping in the forward transformation process. We wish to generalize the work and prove it formally.

It is necessary to show that the concept can be applied to any ER model representing any application domain. On the other hand, a formal proof that can justify the accuracy of the system is an essential goal in Computer Science [17].

The current work aims to show that a one-to-one and onto mapping, as defined in mathematics, exists from an ER model diagram (also called an "ER model") to its RDS. The ER model diagram is created using the modified ER model and transformed to RDS using the modified transformation algorithm. For this purpose, we need to show that a given ER model and its RDS can be expressed as sets.

In the current work, we show that an ER model can be expressed as a set, and the set can be used as a representation of the ER model itself. For this purpose, we use a generic ER model–one that represents phenomena in symbolic notation. A generic ER model can be used as a general representative for exemplifying any ER model from any application domain[13]. We show that the generic ER model can be partitioned into unique segments that each one can represent a meaning in the real world. We call them ER-construct-units and show that such a unit cannot be divided further into smaller units while retaining its meaning. We then show that the set of ER-construct-units of the ER model can be used to represent the ER model itself.

*A. Significance of the research*

The traditional ER model uses conventional graphical constructs to create ER model diagrams. Accordingly, a rectangle is used to represent an entity type, an oval is used to represent an attribute, and a diamond is used to represent a relationship type. The traditional ER model is regarded to be providing a true natural representation of the real world. The model is still popular and widely used for conceptual modelling of databases as well as teaching and learning the database design process. (some recent examples for its use, in practice and research, are: [18-20]).

What we have modified is the traditional ER model. As a result, of the modifications introduced to the traditional ER model and the transformation algorithm, a one-to-one correspondence is established from any ER model diagram created by the modified ER model to the RDS created by the modified transformation algorithm. We argue that, if this modified approach is used, the database designing process will become a much more natural, straightforward, momentary, and trustworthy task for its learners, teachers, and practitioners.

Many automated tools are available for creating ER models for the traditional ER model and its variants. However, no such tool exists to provide a real automatic transformation from the traditional ER model to the RDS. Some tools claiming to be providing an automated transformation can only help the user visualize what he/she is doing with the computer. The user has to transform the ER model diagram to the RDS manually using pointing and clicking devices. The user can monitor and, if necessary, rectify what he/she is doing in the computer. In contrast, we argue that our modified database design approach can provide a high level of and a true nature of automation to the transformation process. Once the ER model diagram is produced, to transform it to the RDS is just a one-click away action. Thus, we believe a Computer-Aided Software Engineering tool (also called CASE tool) could be produced based on our modified approach to automate the transformation process. Tools that are limited to creating ER model diagrams only could also be extended to provid a true automated transformation. We also believe such a CASE tool that we expect will equally enhance the teaching and learning process of database design.

The current research seeks to develop a formal method and use it to validate a systems development method proposed. Thus, we hope the research will contribute significantly to the area of formal methods in software engineering.

*B. Related research*

Kamišalić et al. [21] examine the effectiveness of learning conceptual database design. They found that the manual transformation from a conceptual model to a logical data model can increase students' understanding of the concept. Khaire and Mali [22] presented a web application that can assist in generating an ER diagram automatically. The application needs the user to fill a form it provides to get entities, attributes, and relationships in the application domain as inputs. It then gives the ER diagram as output, automatically[22]. Kuk et al. [23] also present a semi-automated method for generating an ER model from requirements stated in a natural langue. Javed and Lin [24] also undertook a similar study. The method they investigated could generate ER models automatically from requirements stated in a natural language [24].

Yang and Cao, [25] investigated how MySQL Workbench − a visual tool for data modelling − can be used for helping students improve their performance in the ER model to RDS transformation. They also investigated the effects of using MySQL Workbench, in teaching ER to relational transformation. The authors found that visualization of the transformation process could increase the students' interest in it and their engagement with it, as well as their ability to transform the ER model's concepts to the RDS [25]. Wu et al. [26] investigated several versions of the ER model to understand the right ER diagram convention used to teach ER modelling to undergraduate students. Accordingly, they investigated the traditional ER model, the Bachman ER model − the ER model in Bachman notation, and the UML class diagram. The authors found that the traditional ER model is much better than any other model they investigated to introduce ER modeling concepts to students[26].

We will show how our standardized ER to relational transformation process can enhance the above findings. However, the main objective of this paper is to validate formally the standardization that we had undertaken. Thus, with that view in mind, we organize the rest of this paper as follows. In section II, we explore how a real-world small ER model can be partitioned and its ER-construct-units identified. In section III, we deal with a generic ER model and define the ER-construct-units discussed in section II. Section IV extends the work done in section II with a larger ER model. ER-construct-units found in section IV are defined in section V. Section VI presents the conclusion, while section VII details future research.

II. PARTITIONING A REAL-WORLD ER MODEL INTO SEGMENTS

An ER model is a conceptual schema represented as a diagram drawn using ER constructs such as entity types, attributes, and relationship types. It is intended to represent a user application domain in the real world.

On an ER model, the ER constructs do not exist in isolation separated from each other. Still, they exist connected logically as an arrangement that portrays a real-world meaning relevant and vital to the application domain concerned.

For instance, a regular (strong) entity type, including its attributes, is an ER construct arrangement. Fig. 1 shows a regular entity type, which is made up of three ER constructs in such a way that (i) - a primary key(PK) attribute ER construct (Emp_No"), and (ii) - a simple attribute ER construct ("Name") are connected to (iii) - a regular entity type ER construct ("Employee"). The ER model that contains the regular entity type is drawn for representing a portion of a "Company" user application domain.

We argue that the three constructs are the minimum requirement for a regular entity type to be constructed for any application domain, not only for a "Company" application domain. Thus, what is presented in Fig. 1 is the smallest possible regular entity type arrangement. Therefore, it cannot be split or any of its three constructs removed. For instance, if its simple attribute or the PK attribute is removed, the remainder would become meaningless. Hence, each of the three constructs, the PK attribute, the simple attribute, and the regular entity type are mandatory and should exist connected as a single coherent arrangement regardless of the application domain concerned. Therefore, we consider the arrangement to be a single unit of ER constructs.

Even though Fig. 1 regular entity type, which we consider a single unit of ER constructs, cannot be split, it can be expanded by adding one or more simple attributes. For instance, the regular entity type in Fig. 2 expands the regular entity type in Fig. 1 by adding two more simple attributes: "Address" and "Gender." Thus, the regular entity type in Fig. 1 acts as a base and allows other attributes to be added to it. In this context, we consider this single unit of ER constructs to be a base unit of ER constructs. Since it is of a regular entity type, we consider it and call it Regular-entity-base-ER-construct-unit.

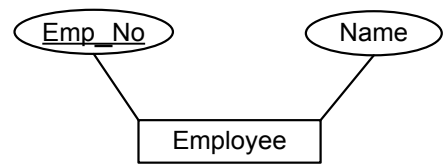

Fig. 1. A regular entity type with two simple attributes

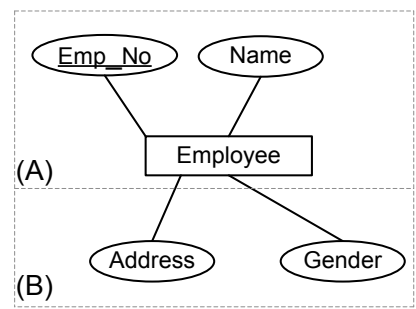

Fig. 2. (A) -The base regular entity type unit, and (B) -the secondary simple attribute unit that are separated

Further, we call the simple attributes that are added to this Regular-entity-base-ER-construct-unit the secondary simple attributes. We call the secondary simple attributes the Regular-entity-secondary-simple-attribute-ER-construct-unit attached to a Regular-entity-base-ER-construct-unit.

Both the Regular-entity-base-ER-construct-unit and the Regular-entity-secondary-simple-attribute-ER-construct-unit are shown partitioned and labelled as (A) and (B), respectively, in Fig. 2. Further, Fig. 2-(B) shows how this Regular-entity-secondary-simple-attribute-ER-construct-unit exists attached to the Regular-entity-base-ER-construct-unit (Fig. 2-(A)).

Next, in section III, we will generalize the concept using a generic ER model proposed by [13].

## III. Partitioning a Small Generic ER Model and Defining its ER-Construct-Units

In the generic ER model [13], the letter "$e$" represents a regular entity type. Consequently, $e_i$ represents the $i^{th}$ regular entity type, where $i \in \mathbb{N} = \{1, 2, 3 \ldots\}$. Further, $k(e_i)$ represents the primary key (PK) attribute. The symbol $s_j(e_i)$ represents the $j^{th}$ simple attribute where, $j \in \mathbb{N}$. Accordingly, the symbols $s_1(e_i)$, $s_2(e_i)$, and $s_3(e_i)$, …, $s_n(e_i)$, represent the $1^{st}$, $2^{nd}$, and $n^{th}$ simple attributes of the entity type $e_i$. The Fig. 3, represents a generic ER-model of this nature. Notice that we reserve the notation, $s_1(e_i)$, to represent the mandatory simple attribute (section II).

In the generic ER model (Fig. 3), the partition named $b(e_i)$ shows the generic equivalent of the Regular-entity-base-ER-construct-unit, the one we showed in the partition (A) in the real-world ER model (Fig. 2)(section II). Accordingly, we formally define the first ER-construct-unit as follows.

*Definition 1:*

In a generic ER model, a regular entity type, $e_i$, its key attribute, $k(e_i)$, and its mandatory simple attribute, $s_1(e_i)$, taken together, is defined as an ER-construct-unit and named as the "Regular-entity-base-ER-construct-unit" and denoted as $b(e_i)$. The unit is shown partitioned and named as $b(e_i)$ in the generic ER model in Fig. 3. Here, the letter "$b$" indicates "base."

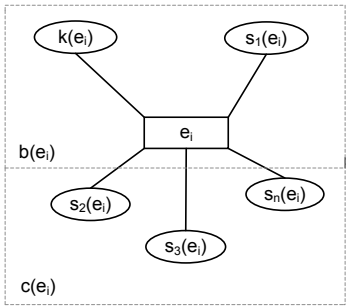

Fig. 3. A generic ER model that represents a regular entity type

The unit is independent and can exist itself meaningfully. It has a semantic meaning itself. The unit acts as a base and lets other constructs to be attached to it.

In the generic ER model (Fig. 3), recall that we reserved the symbol, $s_1(e_i)$ to denote the mandatory simple attribute of the entity type $e_i$. Therefore, we denote a secondary simple attribute by $s_t(e_i)$, where $t \geq 2$. For instance, a set of $n - 1$, where $n > 1$ number of secondary simple attributes of a regular entity type, $e_i$ can be denoted as $s_2(e_i)$, $s_3(e_i)$, …, $s_n(e_i)$.

In the generic ER model (Fig. 3), the partition named $c(e_i)$ shows the generic equivalent of the Regular-entity-secondary-simple-attribute-ER-construct-unit. It is the one we have shown in the partition (B) in the real-world ER model (Fig. 2)(Section II). Accordingly, we define the ER-construct-unit, as follows.

*Definition 2:*

In a generic ER model, the collection of the secondary simple attribute constructs, $\{s_t(e_i) / t \geq 2, t \in \mathbb{N}\}$, connected to a Regular-entity-base-ER-construct-unit, $b(e_i)$ is defined as an ER-construct-unit and named as the "Regular-entity-secondary-simple-attribute-ER-construct-unit" and denoted as $c(e_i)$ (Fig. 3). The unit is shown partitioned and named as $c(e_i)$ in the generic ER model in Fig. 3. The letter "$c$" in $c(e_i)$ indicates the meaning "se*c*ondary." The unit, $c(e_i)$, itself does not provide any semantic meaning when it is taken alone. It provides a meaning only when it is attached to a relevant Regular-entity-base-ER-construct-unit, $b(e_i)$. It always depends on its base unit, $b(e_i)$, for existence.

Fig. 3 shows how a regular entity type, $e_i$, in a generic ER model can be partitioned into two ER-construct-units, named, $b(e_i)$, and $c(e_i)$. It also shows how the two units: $b(e_i)$ and $c(e_i)$, can exist associated with each other and form the segment that consists of the regular entity type, $e_i$ and the attributes connected to it, in a generic ER model. The two units forms a set: $\{ b(e_i), c(e_i) \}$. We assume the set can be used to represent the generic ER model in Fig. 3 that contains the regular entity type, $e_i$.

## IV. Partitioning an ER Model Including a Relationship Type and Identifying its ER-Construct-Units

In this section, we consider an ER model with a relationship type and then identify and partition its ER-construct-units.

Consider the real-world ER model given in Fig. 4 that represents two regular entity types, "Vehicle" and "Project," and a relationship type "AssignedTo" existing in between them. A relationship type like AssignedTo where only two entity types participate in is called a relationship type of degree two. A degree two relationship type like AssignedTo is called a binary relationship type [2]. Notice that in the current work, we only deal with binary relationship types existing in between two different regular entity types. We do not consider recursive relationship types, in the current work.

The ER model in Fig. 4 shows min-max structural constraints on the association of the two entity types with each other via the relationship type. They are shown as two bracketed pairs of values $(min, max)$, as $(m_1, x_1)$ and $(m_2, x_2)$. The pair $(m_1, x_1)$ is placed in between the entity type Vehicle and the relationship type AssignedTo, while $(m_2, x_2)$ is placed in between the entity type Project and the relationship type. We will define and discuss the functionality of the two bracketed (min, max) pairs following how min-max structural constraints have been presented in the literature (e.g., [2])

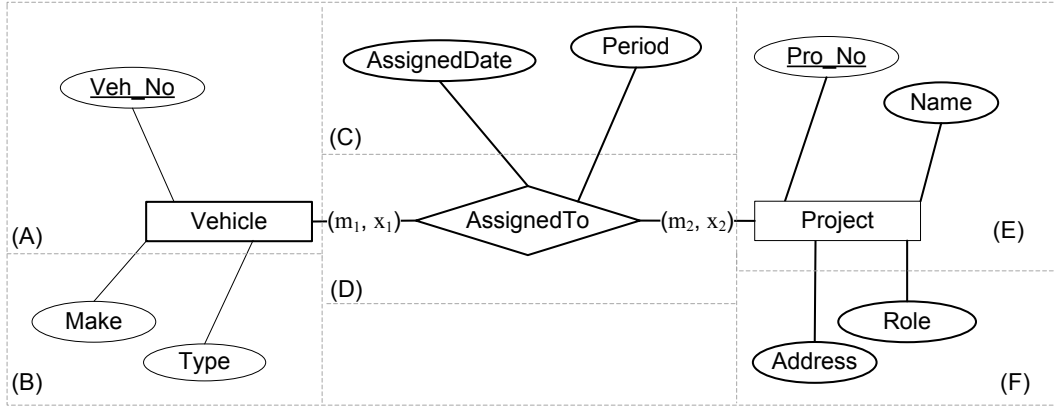

Fig. 4. An ER model that contains a binary one-to-many relationship type and some attributes attached to it

Accordingly, the pair of variables: $m_1$ and $x_1$ lie in the range: $0 \leq m_1 \leq x_1$ and $x_1 \geq 1$, while the pair $m_2$ and $x_2$ lie in the range: $0 \leq m_2 \leq x_2$ and $x_2 \geq 1$. Variables: $m_1$ and $m_2$ represent minimum ($min$) values, while $x_1$ and $x_2$ represent maximum ($max$) values, in their respective ranges. The number $m_1$, in ($m_1, x_1$) means an entity in the entity type Vehicle should participate (via the relationship type AssignedTo) in a minimum $m_1$ number of entities of the entity type Project. The constraint is called the participation constraint. Notice that the number $m_2$ in ($m_2, x_2$) also bears a similar meaning.

On the other hand, the numbers $x_1$ in ($m_1, x_1$) and $x_2$ in ($m_2, x_2$) represent another constraint called cardinality ratio constraint. The constraint is expressed categorizing into three types as one-to-one, one-to-many, and many-to-many, and from one direction of the relationship type to the other.

To understand the participative constraint and the cardinality ratio constraint let us consider the following example (Example 1)– a pair of min-max structural constraints:

[ ($m_1, x_1$) , ($m_2, x_2$) ] $\equiv$ [(0,3) , (1,1)]

Where, $m_1 = 0$, $x_1 = 3$, $m_2 = 1$, $x_2 = 1$.

For instance, $m_1$ represents participation constraint, and $m_1 = 0$ means some entities in the entity type Vehicle may not participate in the relationship type AssignedTo and hence not associate with any entity in the entity type Project. In this case, the participation of the entity type Vehicle in the relationship type AssignedTo is called "partial" or "optional." Similarly, $m_2 = 1$ means every entity in the entity type Project can exist only if it participates in at least one AssignedTo relationship type instance with an entity in the Vehicle entity type. In this case, the participation of the entity type Project in the relationship type AssignedTo is called "total" or "mandatory."

On the other hand, $x_1 = 3$ and $x_2 = 1$ indicate a one-to-many cardinality ratio constraint, which exists in the direction from the entity type Vehicle to the entity type Project. It means an entity in the entity type Vehicle can relate with minimum 0 and maximum 3 entities in the entity type Project, but an entity in the entity type Project can relate with only one entity (maximum) in the entity type Vehicle.

Table 1, below, summarizes two more examples (Example 2 and Example 3) of min-max structural constraints. Example 2 presents a one-to-one cardinality ratio constraint, while Example 3 presents a many-to-many constraint. Notice that Example 1, mentioned above, has already presented a one-to-many constraint.

The binary relationship type consists of the ER constructs: (i)- the relationship type construct "AssignedTo" attached to two regular entity types, "Vehicle" and "Project" and (ii)-a pair of min-max structural constraint constructs denoted by two bracketed pairs of values: ($m_1, x_1$) and ($m_2, x_2$). Each pair is placed on either side of the relationship type.

TABLE I. SUMMARY OF TWO MORE STRUCTURAL CONSTRAINT EXAMPLES

| Participative constraint | | Cardinality ratio constraint | |
|---|---|---|---|
| Example 2 | | | |
| $m_1$ | $m_2$ | $x_1$ | $x_2$ |
| 1 | 0 | 1 | 1 |
| mandatory /total | partial/optional | one-to-one | |
| Example 3 | | | |
| $m_1$ | $m_2$ | $x_1$ | $x_2$ |
| 1 | 2 | 3 | 5 |
| mandatory /total | mandatory /total | many-to-many | |

Assume any of the constructs: (i) or (ii), mentioned above, does not exist in the structure. Then the relationship type may not exist, and the remainder may become meaningless. Therefore, for a meaningful relationship type to exist, both constructs must exist with binding together and acting as a single unit.

Two simple attributes: "AssignedDate" and "Period" are attached to the relationship type AssignedTo in Fig. 4. They are optional attributes. That is, they may or may not exist.

Thus, we consider the relationship type consisting of the relationship type construct and the min-max structural constraint construct to be a separate ER-construct-unit.

Since the attributes can sometimes exist attached to the relationship type, the relationship type acts as a base and allows other constructs (attributes) to be attached to it. In this context, we deem the relationship type to be a base ER-construct-unit.

The relationship type exists attached to two Regular-entity-base-ER-construct-units. If the two Regular-entity-base-ER-construct-units do not exist, the relationship type does not exist. Thus, the relationship type is a dependent unit that depends on the two Regular-entity-base-ER-construct-units. Accordingly, the relationship type ER-construct-unit depends on the Regular-entity-base-ER-construct-units for its existence. In the meantime, it acts as a base and allows other constructs (attributes) to be attached to it.

We name the relationship type to be a Binary-relationship-type-ER-construct-unit. Notice that the unit is separated and highlighted by a dashed line and labelled as (D) in Fig. 4.

The attributes attached to the relationship are optional. That is, they may or may not exist attached to the relationship type. Even if they exist, the number of them varies. Thus, the simple attributes attached to the relationship type seems to have a particular behavior inherent to them. Therefore, we consider the simple attributes attached to a Binary-relationship-type-ER-construct-unit to be a separate ER-construct-unit. We call the unit a Simple-optional-attribute-ER-construct-unit attached to a Binary-relationship-type-ER-construct-unit. Notice that this unit is separated by a dashed line and labelled as (C), on the ER schema, in Fig. 4.

The generic equivalents of the ER-construct-units: (C) and (D) in Fig. 4 will be defined in the next section.

## V. PARTITIONING A MODERATE LEVEL GENERIC ER MODEL AND DEFINING ITS ER CONSTRUCT UNITS

For this purpose, we again use the generic ER model proposed by [13]. The generic ER model uses the symbol, $r_v(e_i, e_j)$, where $v \in \mathbb{N}$, for denoting a binary relationship type existing between two regular entity types $e_j$ and $e_j$. Attributes attached to the relationship type are denoted as $s_1(r_v(e_i, e_j))$, $s_2(r_v(e_i, e_j))$,..., $s_t(r_v(e_i, e_j))$, where $t \in \mathbb{N}$. The min-max values are denoted as variables: $m_1$, $x_1$, $m_2$,..and $x_2$ Fig. 5 shows a binary relationship type existing in a generic ER model.

In the generic ER model (Fig. 5), the partition named $b(r_v(e_i, e_j))$ shows the generic equivalent of the Binary-relationship-type-ER-construct-unit, which we have shown in the partition (D) in the real-world ER model (Fig. 4). Accordingly, we formally define the ER-construct-unit as follows.

Definition 3:

In a generic ER model, the arrangement that consists of the two ER constructs: (i) − a relationship type construct, $r_v(e_i, e_j)$, which is attached to two regular entity base ER construct units, $b(e_i)$ and $b(e_j)$, and (ii) − a min-max structural constraint construct denoted by two bracketed pairs of values: $(m_1, x_1)$ and $(m_2, x_2)$ where each bracketed pair is placed on either side of the relationship type, is defined to be an ER-construct-unit. The unit is named as the Binary-relationship-type-ER-construct-unit and denoted as $b(r_v(e_i, e_j))$. The unit is shown partitioned and named as $b(r_v(e_i, e_j))$ in the ER model in Fig. 5. The letter "$b$" indicates the meaning "base".

Notice that depending on the actual numerical values of the min-max variables, the relationship type may get either of the forms: one-to-one, one-to-many, or many-to-many. However, the constitution and the shape of the ER-construct-unit are not to be changed for any form of the relationship type: one-to-one, one-to-many, or many–to-many.

In the generic ER model (Fig. 5), the partition named - $p(r_v(e_i, e_j))$ shows the generic equivalent of the Simple-optional-attribute-ER-construct-unit, the one we have shown in the partition (C) in the real-world ER model (Fig. 4). Accordingly, we formally define the ER-construct-unit as follows.

*Definition 4:*

In a generic ER model, the collection of the simple attributes attached to a Binary-relationship-type-ER-construct-unit, $b(r_v(e_i, e_j))$, is defined to be an ER construct unit. The unit is named as the Simple-optional-attribute-ER-construct-unit attached to Binary-relationship-type-ER-construct-unit, $b(r_v(e_i, e_j))$. The unit is partitioned and denoted as $p(r_v(e_i, e_j))$ in the ER model in Fig. 5. The letter "$p$" represents the meaning "optional". The unit is an optional unit, that is, it may or may not exist attached to a unit, $b(r_v(e_i, e_j))$. If it exists, its number of attributes may vary.

Accordingly, Fig. 5 shows how a Binary-relationship type, $r_v(e_i, e_j)$), in a generic ER model can be partitioned into two ER-construct-units, named, $b(r_v(e_i, e_j))$, and $p(r_v(e_i, e_j))$. It also shows how the two units: $b(r_v(e_i, e_j))$ and $p(r_v(e_i, e_j))$ can exist associated with each other and form the relationship type, $r_v(e_i, e_j)$) in a generic ER model.

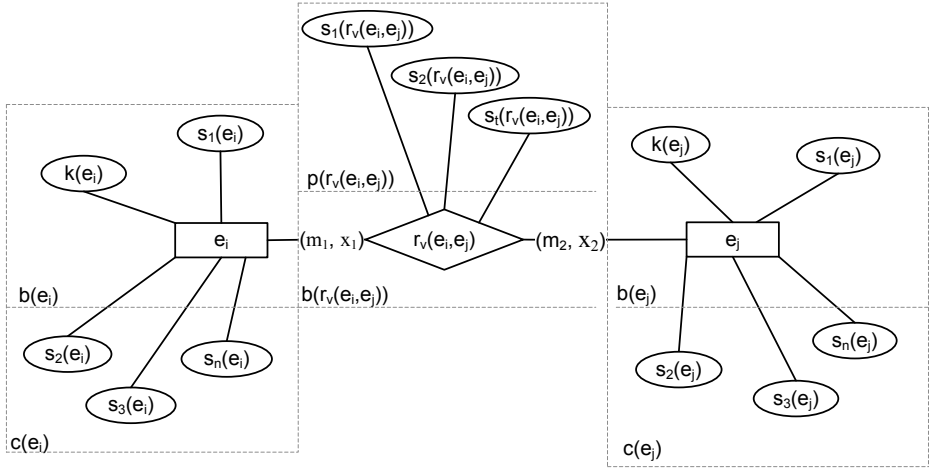

Fig. 5. A generic ER model containing a binary one-to-many relationship type attached to two regular entity types

## VI. Conclusion

We have shown (section III) that the regular entity type, $e_i$, in the ER model (Fig. 3) can be partitioned into two distinct ER construct units, $b(e_i)$ and $c(e_i)$. The same partitions and the ER construct units: $b(e_i)$ and $c(e_i)$ exist in the generic ER model in Fig. 5. Similarly, the regular entity type, $e_j$, in the generic ER model (Fig. 5) can also be partitioned into two ER-construct-units, $b(e_j)$ and $c(e_j)$. We also showed that the binary-relationship type, $r_v(e_i, e_j))$, in the generic ER model (Fig. 5) can be partitioned into two ER-construct-units, $b(r_v(e_i, e_j))$, and $p(r_v(e_i, e_j))$.

Accordingly, the entire generic ER model in Fig. 5 can be partitioned into six ER-construct-units, namely, $b(e_i)$, $c(e_i)$, $b(r_v(e_i, e_j))$, $p(r_v(e_i, e_j))$, $b(e_j)$, and $c(e_j)$. The six partitions are distinct: that is, any one of them does not overlap or penetrate into another. They all together cover the entire generic ER model (Fig. 5).

The six distinct ER-construct-units form a set: $\{b(e_i), c(e_i), b(r_v(e_i, e_j)), p(r_v(e_i, e_j)), b(e_j), c(e_j)\}$. We assume that the set can be used to represent the generic ER model (Fig. 5).

On the other hand, a generic ER model can represent any real-world ER model[13]. Thus, we conclude that any real-world ER model that contains a binary relationship type that exists between two regular entity types can be viewed as a set of six elements and the set can be used as a representation of the ER model.

## VII. Future research Imerging from the current research

The current paper presents a part of an ongoing reach. Its results will be used in the future for further research expected. Accordingly, in future research, we will transform the moderate level generic ER model (Fig. 5) to a relational database schema (RDS). We will use the modified transformation algorithm for this purpose. We will then partition the RDS into segments, which we call Relation-schema-units. Next, we show that a mapping that is one-to-one and onto exists from the set representing the generic ER model to the set representing its RDS. We will then show that the information represented on the ER model is preserved on the RDS.

## VIII. Implications of the Research series

We argued that a one-to-one and onto correspondence from the ER model to the RDS not only preserve information from the ER model to the RDS. It also should be a basis for automating the transformation process from the ER model to the relational model. In section 1, we stated that a CASE tool can be created for automating the process.

We believe the CASE tool we expect can extend the work of Khaire and Mali [22]. The tool can be integrated with the web application that they have proposed. The CASE tool can then be used to automatically transform an ER model produced by the web application to the relational model. The CASE tool we expect should be able to be integrated with any other CASE tool that creates ER models (e.g. ERDplus - https://erdplus.com/ ) to transform them to the RDS automatically. Further, a CASE tool we expect also can extend the works of [23], and [24] (Section 1), in the same manner, mentioned above.

Going beyond the visualization of a computer-aided transformation process proposed by Yang and Cao [25], the CASE tool we expect could undertake the entire transformations process and perform it purely automatically without letting a user be intervened at intermediate stages for making adjustments. Even if the traditional ER model is claimed to be more suitable for teaching ER modeling concepts [26], in our view, the database designing process cannot be limited to just ER modeling only. Once an ER model diagram is created, it needs to be transformed to the RDS. The created RDS should be accurate and a one that preserves the information of its predecessor ER model. Without obtaining the skill that how an ER model can be transformed to the RDS, accurately and with preserving information, the database design and learning process is deemed to be incompleted. We argue that our modified approach comprising the ER model and the transformation algorithm that we have modified can fill this gap. It provides a hassle-free learning process. The reason the ER modeling and transformation rules are now apparent, straightforward, and ambiguous free. They provide a one-to-one transformation from the ER model to the RDS, which will also automate the transformation process. An automated tool can help students to validate their manual transformations and iteratively improve them until a correct RDS is reached as the output. The same advantage is equally applicable to the practitioners as they no longer need worrying about how models can be transformed from one to the other from the ER model to the RDS. A CASE tool will do the job for them.

Except for our ongoing researches for formal validation of our approach, empirical researches can be undertaken with learners, teachers, and practitioners aiming to assess our claims about the impact of the approach on improving the efficiency and productivity of them. If a CASE tool is produced, it can also be used as a tool for empirical validation of the approach.